\documentclass[%
 reprint,
superscriptaddress,
 amsmath,amssymb,
 aps,
pra,
]{revtex4-2}

\usepackage{graphicx}% Include figure files
\usepackage{dcolumn}% Align table columns on decimal point
\usepackage{bm}% bold math
\usepackage{amsfonts}
\usepackage{amsmath}
\usepackage{amsthm}
\usepackage{physics}
\usepackage{caption}
\usepackage{subcaption}

\newcommand{\Z}{\mathbb{Z}}
\newcommand{\R}{\mathbb{R}}
\renewcommand{\L}{\Lambda}
\newcommand{\B}{B}
\renewcommand{\b}{\bm{b}}
\newcommand{\vol}{\text{vol}}
\newcommand{\z}{\bm{z}}
\newcommand{\x}{\bm{x}}
\newcommand{\G}{\bm{G}}
\renewcommand{\v}{\bm{v}}
\newcommand{\Q}{\hat{Q}}
\newcommand{\pZ}{\hat{Z}}
\newcommand{\I}{\mathbb{I}}
\renewcommand{\O}{\mathcal{O}}
\renewcommand{\d}{\mathrm{d}}

\newtheorem{definition}{Definition}
\newtheorem{theorem}{Theorem}

\newtheorem{lemma}{Lemma}

\begin{document}

\preprint{APS/123-QED}

\title{Quantum algorithmic solutions to the shortest vector problem on simulated coherent Ising machines}

\author{Edmund Dable-Heath}
\affiliation{Electronic and Electrical Engineering Department, Imperial College London, Prince Consort Road, SW7 2AZ, United Kingdom}
\affiliation{The Applied Research Centre, The Alan Turing Institute, British Library, 96 Euston Rd., London NW1 2DB}
\author{Laura Casas}
\affiliation{Physics Department, Blackett Laboratory, Imperial College London, Prince Consort Road, SW7 2AZ, United Kingdom}
\author{Victor Hertz}
\affiliation{Electronic and Electrical Engineering Department, Imperial College London, Prince Consort Road, SW7 2AZ, United Kingdom}
\author{Christian Porter}
\affiliation{Electronic and Electrical Engineering Department, Imperial College London, Prince Consort Road, SW7 2AZ, United Kingdom}
\author{Florian Mintert}
\affiliation{Physics Department, Blackett Laboratory, Imperial College London, Prince Consort Road, SW7 2AZ, United Kingdom}
\affiliation{Helmholtz-Zentrum Dresden-Rossendorf, Bautzner Landstraße 400, 01328 Dresden, Germany}
\author{Cong Ling}
\affiliation{Electronic and Electrical Engineering Department, Imperial College London, Prince Consort Road, SW7 2AZ, United Kingdom}

\date{\today}% It is always \today, today,

\begin{abstract}
    Quantum computing poses a threat to contemporary cryptosystems, with advances to a state in which it will cause problems predicted for the next few decades. Many of the proposed cryptosystems designed to be quantum-secure are based on the Shortest Vector Problem and related problems. In this paper we use the Quadratic Unconstrained Binary Optimisation formulation of the Shortest Vector Problem implemented as a quantum Ising model on a simulated Coherent Ising Machine, showing progress towards solving SVP for three variants of the algorithm.
% \begin{description}
% Secondary publications and information retrieval purposes.
% You may use the \texttt{description} environment to structure your abstract;
% use the optional argument of the \verb+\item+ command to give the category of each item. 
% \end{description}
\end{abstract}

\maketitle

%\tableofcontents

\section{\label{sec:intro}Introduction}
    With the development on scalable quantum computing on the horizon \cite{almudever2018towards, sete2016functional, o2007optical} the effort to create the quantum-secure cryptographic systems for classical computation has developed a clear front runner in lattice based cryptography (LBC) \cite{moody2021nist}. Whilst cryptosystems based on lattices appear to be resistant to related gate model quantum attacks that have challenged RSA \cite{milanov2009rsa, shor1999polynomial}, amongst other contemporary systems \cite{kumar2021state}, there is an ongoing endeavour to assess their security under a variety of quantum computational models. The gate model of quantum computing can be thought of as the quantum equivalent to the standard digital model of classical computation, however there are many alternative models that draw closer comparison to analogue computing. In particular, given the common practice of formulating computational problems as Ising models \cite{lucas2014ising}, systems such as Coherent Ising Machines (CIM) \cite{wang2013coherent} that implement a physical Ising model could provide efficient methods for solving these problems. Here we investigate how CIMs could be used to question the security assumptions of lattice based cryptography.

    % \subsubsection{Post-Quantum Cryptography}
    The advent of quantum computing - especially Shor's algorithm \cite{shor1999polynomial} - has led to a need to question whether the current cyptographic standards are good enough. Whilst effective direct attacks on cryptographic systems by quantum computers are a little way off \cite{ugwuishiwu2020overview} this has led to the National Institute of Standards and Technology (NIST) to run a standardisation competition \cite{moody20192nd}. 

    Weakness to quantum attacks comes from attacks on a piece of the fundamental design of cryptographic systems: the obfuscation of information by application of a computationally hard mathemtical problem. Specifically problems that exhibit a trapdoor - easy to compute functions with an inverse that is hard to compute without already possessing the answer. A key example of this is the factorisation of semi prime number $ n=pq $ into their unique prime factors $ p $ and $ q $. Knowing one or both of the factors solves the problem simply, however without this one must resort to computationally expensive factorisation algorithms. RSA \cite{milanov2009rsa} bases its security on this problem, with Diffie-Hellman \cite{maurer2000diffie} and El-Gamal \cite{meier2005elgamal} based on related problems. It is this exact problem (and related problems), however, that Shor's algorithm purports to solve in a computationally efficient manner. 

    Proposal schemes to the NIST standardisation procedure look to replace this underlying hard problem with problems that can't be efficiently computed by quantum computational models. Several of the proposed PQC cryptosystems are based in LBC: learning-with-errors (LWE) \cite{regev2010learning} and its structured adaptations ring-LWE \cite{lyubashevsky2010ideal} and module-LWE \cite{boudgoust2023hardness}, NTRU \cite{hoffstein1998ntru}.

    The computational problems in LBC center on minimising the distance between objects in high dimensional spaces. The Shortest Vector Problem (SVP) poses finding the closest non-zero vector to the origin, picking a different target vector in the ambient space adapts this to the Closest Vector Problem.

    Coherent Ising Machines use quantum optical effects to implement a physical Ising model, allowing one to program the solution to a computational problem into the energy eigenstates, for which the device optimises for. Recent physical implementations of these devices have seen up to 100,000 spin sites \cite{honjo2021100}, with efficient operation in solving several standard performance heuristic problems, such as the Max-Cut problem \cite{haribara2016coherent}. However, the assessment of the performance of these devices on real-world problems is still in its early days. In this paper we develop an algorithm that applies a CIM as an SVP solver, testing its performance under several key metrics for a low dimensional set of lattices. Key to the efficient running of any computational system is effective use of the resources at hand, for which we also present several variants of the CIM SVP solver and associated spatial complexity bounds. 
    
    The paper is structured as follows: Section \ref{sec:prelims} lays out some preliminaries. In section \ref{sec:QUBO} the quadratic unconstrained binary optimisation (QUBO) form of the shortest vector is introduced and the quantum Ising model encoding variants given. Within this section we also outline our method for removing the zero vector of the lattice from the search space to ensure that the shortest non zero vector is the ground state. The implementation of the QUBO SVP into an Ising model using various encoding methods is shown with complexity bound on the number of qubits required for each encoding derived. Implementations of the QUBO SVP on a simulated CIM are shown in section \ref{sec:sims}. Simulations include a low-dimensional implementation of the three qubit encodings presented in section \ref{sec:QUBO}, and implementation of the constraint to remove the zero vector as the ground state.

\section{\label{sec:prelims}Preliminaries}
    Some general notes on notation: vectors are notated by lowercase boldface $ \bm{x} $, matrices by uppercase boldface $ \bm{B} $ and lattices by $ \L $. All norms $ \|\bm{x}\| $ refer to the 2-norm: \[ \|\bm{x}\|_2 = \sqrt{\sum_i x_i^2}. \]
    Operators are denoted $ \hat{A} $, with $ \hat{H} $ referring to the Hamiltonian operator. Where it is clear from context the tensor product $ \bigotimes $ is dropped.

    \subsection{\label{sec:latt_probs}Lattices}
\begin{figure}
            \centering
            \includegraphics[scale=0.17]{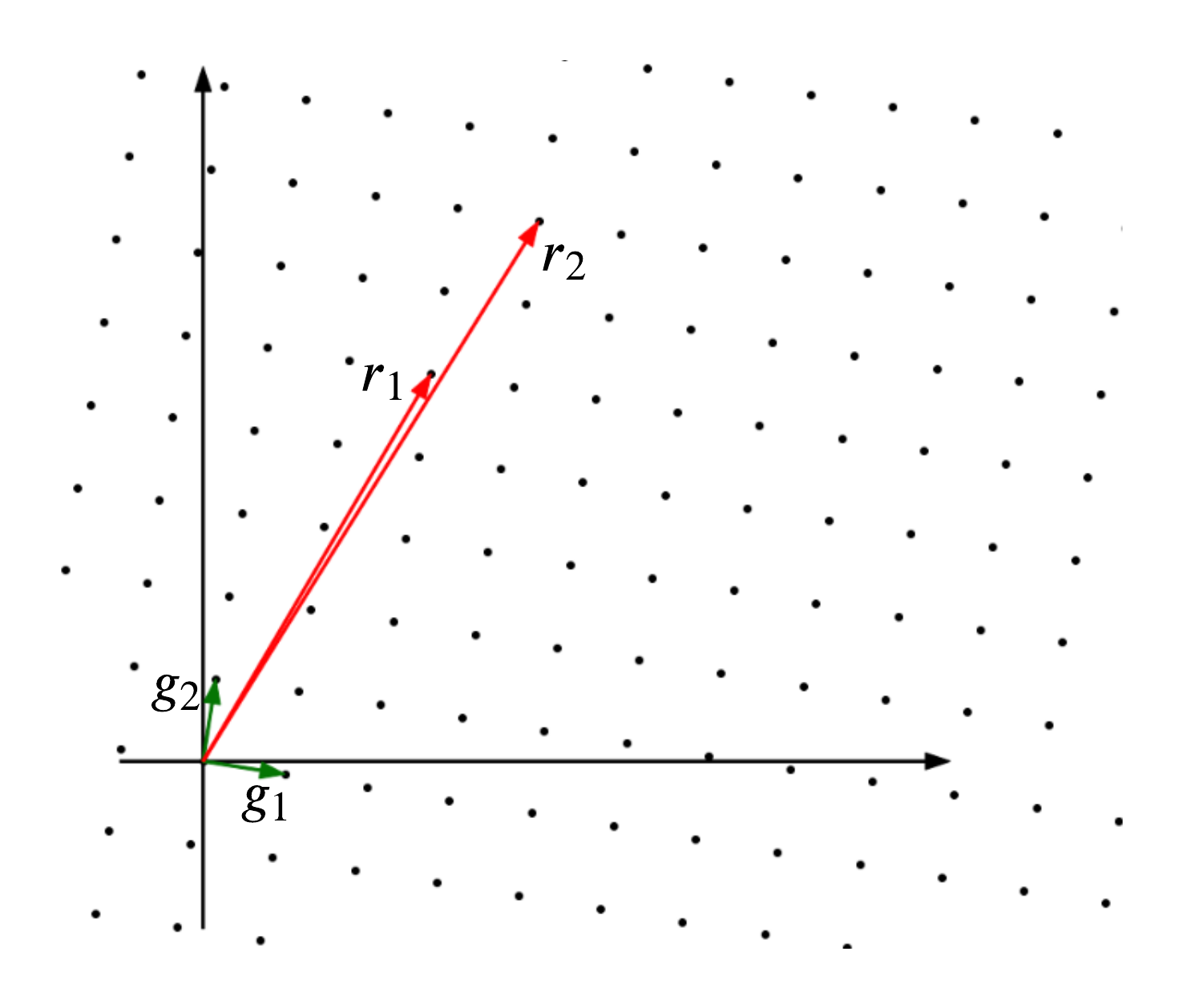}
            \caption{A two dimensional example lattice (here a rotation $ \Z^2 $). Two example bases are pictured: the green vectors, $ \{g_1,g_2\} $, represent a `good' basis with short, orthogonal vectors, and the red vectors, $ \{r_1,r_2\} $, a `bad' basis with long, non-orthogonal vectors.}
            \label{fig:2d_latt}
        \end{figure}

        Lattices are discrete subsets of $ \R^n $ with a regular, repeating structure. See Fig. \ref{fig:2d_latt} for a two dimensional example. They form linear spaces containing the origin, making them attractive for computational problems. 

        \begin{definition}
            A \emph{lattice} $ \L\subset\R^n $ is a discrete additive subgroup of $ \R^n $. That is $ \bm{0}\in\L $, if $ \bm{x}, \bm{y}\in\L $ then $ -\bm{x},~\bm{x}+\bm{y}\in\L $ and for any $ \bm{x}\in\L $ there is a neighbourhood around $ \bm{x} $ in $ \R^n $ which contains no other lattice points.
        \end{definition}

        A lattice is often defined by a basis $ \B=\{\b_1,\ldots,\b_k\} $, with the lattice generated by all integer combinations of the basis vectors,
        \begin{equation}\label{eq: basis definition}
            \L(\B\Z^n) = \left.\left\{\sum_{i=1}^k z_i\b_i\right| z_i\in\Z\right\}.
        \end{equation}
        Bases $ \B $ are not unique as any two bases can be related by a unimodular transform. The number of basis vectors $ k $ describes the rank of the lattice, with a lattice $ \L\subset\R^n $such that $ k=n $ known as full rank. In this work we restrict ourselves to full rank lattices. Any basis can be described as being of good or bad quality with better bases having shorter and closer to orthogonal vectors.

        The following definitions, \ref{def:dual latt} - \ref{def:well rounded}, are key for theorems presented in section \ref{sec:QUBO}.

        \begin{definition}[Dual Lattice] \label{def:dual latt}
            The \emph{dual lattice} $ \L^* $ of a lattice $ \L(\B) $ is defined as
            \begin{equation}
                \L^*(\B) = \{\bm{x}\in\R^n\mid\langle\bm{x},\L\rangle\subset\Z\}.
            \end{equation}
        \end{definition}

        \begin{definition}
            The \emph{fundamental parallelpiped} $ \mathcal{P}(\B) $ is defined as the set of point in $ \R^n $
            \begin{equation}
                \mathcal{P}(\B) = \{\bm{x}\cdot\B \mid \bm{x}\in[0,1)^n\},
            \end{equation}
            the volume of which is known as the lattice \emph{covolume} and denoted $ \vol(\L) $.
        \end{definition}
        The geometry of the fundamental parallelpiped depends on the basis, however the covolume is a lattice invariant and independent of the choice of basis. One can compute the covolume as $ |\det(\B)| $. One can further characterise the geometry of the basis by how far the vectors are from orthonormal using the orthogonality defect.
        \begin{definition} [Orthogonality defect] \label{def:ortho defect}
            Given a basis for a lattice $ \L(\B) $ the orthogonality defect of the basis is defined as
            \begin{equation}
                \delta(\B) = \left(\frac{\prod_{i=1}^n\|\b_i\|}{\vol(\L)}\right)^2.
            \end{equation}
        \end{definition} 

        Given a basis for a lattice in which one wishes to solve computational problems one can improve the quality of the basis - a process known as basis reduction - prior to the application of any algorithms. The most widely used reduction method is LLL reduction \cite{nguyen2010lll} due to its polynomial time complexity. Another important reduction is HKZ reduction \cite{hanrot2011analyzing}, which has an exponential time complexity, though returns higher quality bases that LLL. The exact definitions of these is not necessary for this paper, however qualities of bases reduced by both methods are discussed in section \ref{sec:QUBO}.

        The lattice problem that is primarily of interest in this work is the \emph{shortest vector problem}:
        \begin{definition}[The Shortest Vector Problem (SVP)]
            Define $ \lambda_1(\L) $ to be the length of the shortest non-zero vector in $ \L $. Then the \emph{shortest vector problem (SVP)} is: given a basis $ \B $ of a lattice $ \L $, compute a vector $ \bm{v}\in\L $ such that $ \|\bm{v}\| = \lambda_1(\L) $.
        \end{definition}
        In every lattice there are at least two shortest vectors, $ \bf{v}_1 $ and $ -\bf{v}_1 $. The sequence of vector norms with increasing length are known as the sucessive minima, and denoted $ \lambda_i $ for the $ i^{th} $ minima. Approximate versions of the above problem and related lattice problems - in which one is looking to compute short vectors typically with length within a multiple of the shortest - are also an important class of lattice problem. The shortest vector problem forms the security basis for many post-quantum cryptosystems, due to a reduction from SVP to the LWE problem \cite{albrecht2014efficacy}.

        Finally we note two important classes of lattices, cyclic and negacylic lattices. These lattices are themselves a subset of ideal lattices, lattices based on the correspondence between ideals of polynomial rings and integer sublattices $ \L\subset\Z^n $. Cyclic lattices are ideal lattices constructed by applying a rotational shift operator to a vector formed by the coefficients of a polyonimal to form successive rows of a basis matrix, with negacyclic lattices formed by a rotational shift operator that introduces a negative sign to certain coefficients on each application. Negacyclic lattices form the basis for cryptosystems such as ring-LWE \cite{lyubashevsky2010ideal} and Falcon \cite{fouque2018falcon}, providing greater efficiency than their unstructured counterparts, at the cost of a reduction in security. The full technical definition of such lattices is not necessary, however we take advantage of two key properties of cyclic and negacylic lattices for the complexity proofs of section \ref{sec:QUBO}: the dual of a cyclic or negacyclic lattice is also a cyclic or negacylic lattice, and that these lattices are examples of well rounded lattices. The latter property is formally defined as follows,
        \begin{definition}[Well Rounded Lattice]\label{def:well rounded}
            Let $ \L\subset\R^n $ be a full rank lattice, and define the set of minimal vectors of the lattice as,
            \begin{equation}
                S(\L) := \{\v\in\L\mid\|\v\|=\lambda_1\},
            \end{equation}
            where $ \lambda_1 $ is the first successive minima of the lattice. Then we see that $ \L $ is \emph{well rounded} if $ \emph{Sp}_{\R}(S(\L))=\R^n $.
        \end{definition}
        In other words the set of vectors in a well rounded lattice solving the shortest vector problem span $ \R^n $, meaning there are at least $ n $ shortest vectors in such lattices, all sitting on a sphere of radius $ \lambda_1 $ centered at the origin.

    \subsection{Computational Ising Models}
        Ising models have long since moved beyond exclusively modelling ferromagnetic materials and now provide a convenient computational framework for problems that can be expressed as a minimisation over a set of binary variables \cite{tanaka2017quantum, lucas2014ising}. In a quantum setting they form the basis for adiabatic computation \cite{albash2015reexamining} and coherent Ising machines. In our case that they have a convenient resemblance to quadratic unconstrained binary optimisation problems \cite{lewis2017quadratic}. This is expanded upon in section \ref{sec:QUBO}.

        Typically one considers particles with spin (usually two level spin) on a lattice graph (which can be generalised to arbitrary graph topology). Particles at graph nodes in this model have interactions with other sites for which they share an edge, and an overall external magnetic field applied across all sites. For a classical Ising model one considers a vector of spins - the spin configuration $ \bm{s} = (s_k)_{k\in\L} $ where $ s_k\in\{-1,1\} $. Each pair of adjacent sites $ i, j\in\L $ has an interaction term $ J_{ij} $ and an external magnetic field term $ h_j $, allowing the energy of any particular configuration to be given by the Hamiltonian,
        \begin{equation}\label{eq: Classical Ising Hamiltonian}
            H(\bm{s}) = -\sum_{\langle i,j\rangle} J_{ij}s_i s_j - \sum_j h_j s_j.
        \end{equation}
        To map this to the quantum domain one replaces the spin variables $ s_k $ with Pauli-$ z $ operators $ \hat{Z}_k $ acting on the $ k^{th} $ qubit of the system. The Pauli operators (along with the identity) form a basis for qubit operators and can be expressed as a set of four $ 2\times2 $ matrices:
        \begin{align}
            &\I = \begin{pmatrix}
                1 & 0 \\
                0 & 1
            \end{pmatrix}, &&\hat{X} = \begin{pmatrix}
                0 & 1 \\
                1 & 0
            \end{pmatrix},\\
            &\hat{Y} = \begin{pmatrix}
                0 & -i \\
                i & 0
            \end{pmatrix}, &&\hat{Z} = \begin{pmatrix}
                1 & 0\\
                0 & -1
            \end{pmatrix}.
        \end{align}
        
        By combining the identity operator and the Pauli-$ z $ operator one can map binary variables in the computational basis with eigenvalues $ \{-1, +1\} $ to $ \{0, 1\} $ using the term $ \frac{\I+\hat{Z}}{2} $ within an Ising Hamiltonian,
        \begin{equation}\label{eq: Quantum Ising Ham}
            \hat{H} = \sum_{\langle i,j\rangle}J_{ij}\frac{\I_i+\pZ_j}{2}\bigotimes\frac{\I_j+\pZ_j}{2} + \sum_{i}h_i\frac{\I_i+\pZ_i}{2}.
        \end{equation}

        Several architectures now exist that allow a programmer to control the values of the coefficients of the couplings $ J_{ij} $ and the external magnetic field $ h_i $. These values can be set such that the ground state - or related low energy states - encode solutions to the computational problem they wish to solve.

    \subsection{Error Correcting Coherent Ising Machines}
        Coherent Ising Machines (CIM) have seen a lot of progress in the last decade, with advances in the physical CIMs \cite{yamamoto2017coherent, honjo2021100} and success in many performance heuristic problems in both simulation and physical CIMs \cite{hamerly2019experimental, haribara2016coherent}. They physically implement an Ising model by storing the spins as wave packets generated by optical parametric oscillators (OPO) and passing them through optical circuits to implement interactions. This allows them to solve computation problems that can be expressed as Ising models. At current the CIMs are restricted to binary variables and lack an external field component, however have the ability to implement any graph topology required for the Ising system considered.

        The standard coupling scheme performed by CIMs suffers from amplitude heterogenity, meaning that the programmed Ising Hamiltonian is incorrectly mapped to the system resulting in unsuccessful operations \cite{wang2013coherent}. This problem can be resolved by the application of error correction within the CIM. Within this work we use error correction by chaotic feed back control (CIM-CFC) \cite{kako2020coherent}. Alongside the spin variables $ x_i $, auxiliary error variables $ e_i $ are introduced, with the following deterministic equations emulated on a CIM,
        \begin{align}
            z_i &= e_i\xi\sum_jJ_{ij}x_j, \\
            \dfrac{\d e_i}{\d t} &= -\beta e_i(z_i^2 - \alpha), \\
            \dfrac{\d x_i}{\d t} &= -x_i^3 + (p-1)x_i - z_i. 
        \end{align}
        Here $ J_{ij} $ are the coupling constants for the Ising system, $ \alpha,~\beta $ and $ p $ are system parameters and $ \xi $ is a normalisation constant for $ J_{ij} $. Typically one can achieve greater success by varying the parameters, allowing the system to explore the search space rapidly before converging to a solution. Parameter choices are discussed in section \ref{sec:sims}. As an Ising solver one considers the spin configuration $ \sigma_i = \text{sign}(x_i) $ as a possible solution to the associated Ising problem defined by $ J_{ij} $. In simulation Gaussian noise terms are often included, however to achieve a diverse set of spin trajectories within this work we consider random initial spin amplitudes $ x_i $.

\section{\label{sec:QUBO}QUBO Formulation and Spatial Complexity Bounds}
    When implementing the SVP as a computational problem it is convenient to consider each integer coefficient with respect to the basis vectors as a variable, as defined by Eq. \eqref{eq: basis definition}, for a lattice $ \L=\L(\B)\subset\R^n $. The resultant integer vector uniquely determines a lattice vector whose length can easily be calculated using the basis matrix. This approach also allows us to accurately limit the binary variables required whilst ensuring that the shortest vector is still within the solution space - an especially important feature for NISQ era architectures. There is a natural correspondence between quadratic unconstrained binary (QUBO) forms with a general cost,
    \begin{equation}
        C(s_1 s_2 \ldots s_n) = c + \sum_{i\neq j}c_{ij}s_i s_j + \sum_{i}c_{ii} s_i,
    \end{equation}
    where $ s_1 s_2\cdots s_n $ is a string of binary variables with $ s_i\in\{0,1\} $ and $ c, \{c_{ij}\}_{1\leq i,j\leq n} $ coefficients, and the Ising Hamiltonian, Eq. \eqref{eq: Quantum Ising Ham} \cite{joseph2020not, joseph2021two, albrecht2022variational}. Further to this, if all of the variables are integers then within the solution space exists the vector in which all the coefficients are zero, the trivial answer to which vector is the shortest within a lattice. By adapting the QUBO formulation it is possible to impose the condition that the solution space, $ \mathcal{S} $, becomes $ \mathcal{S}\subset\L\backslash\bf{0} $ \cite{albrecht2022variational}.
    
     First we examine the QUBO formulation, followed by variants on the integer encoding method. Then we present a method for imposing the removal of the zero vector from the solution space. Finally several spatial complexity theorems are given.

    \subsection{The QUBO Formulation of SVP}
        Given $ \bm{v}\in\L(\B)\subset\R^n $ - where the rows of a matrix $ \B $ span the lattice - $ \bm{v} $ can be expressed as $ \bm{v}=\bm{z}\B $ for $ \bm{z}\in\Z^n $. The length of said vector can therefore be expressed as $ \|\v\|^2 = \bm{z}\B\B^T\bm{z}^T = \bm{z}\G\bm{z}^T $ where $ \G=\B\B^T $ is the Gramm matrix. This allows for a reformulation of the shortest vector problem as a quadratic unconstrained integer optimisation problem,
        \begin{align}\label{eq: QUBO form}
            \lambda_1(\L)^2 &= \min_{\v\in\L\backslash\{\bm{0}\}}\|\v\|^2 \nonumber \\
            &= \min_{\z\in\Z^n\backslash\{\bm{0}\}}\left(\sum_{i=1}^nz_i^2\G_{ii} + 2\sum_{1\leq i<j\leq n}z_iz_j\G_{ij}\right).
        \end{align}
    
\begin{figure}
            \centering
            \includegraphics[scale=0.4]{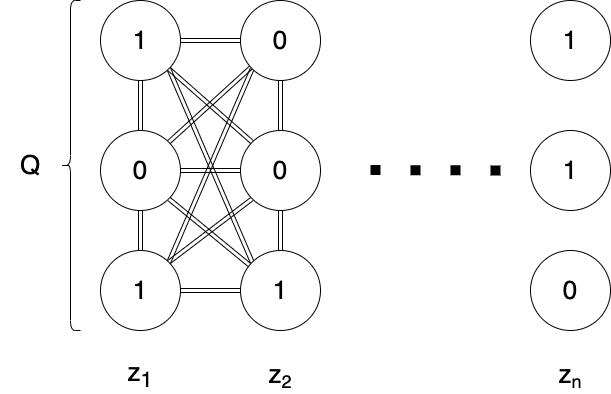}
            \caption{Schematic diagram for encoding multiple qubits as qudits representing integer coefficients for lattice vectors. Each column in this diagram represents a register of qubits that encode a integer decoded by $ Q $ to produce the integer coefficient for each dimension $ z_i $. The links here represent the nearest neighbours, in the QUBO case we require a complete graph topology.}
            \label{fig:qubit_diag}
        \end{figure}

        In the above formulation the variables $ z_i\in\bm{z} $ are assumed to have integer values, a quality unfortunately not accessible by the majority of contemporary quantum hardware, including the CIM. Ideally we would have access to a system composed of qudits, in which the spin level $ d $ could be tuned to the integer range we require. In lieu of such a system one must instead encode integer values into registers of qubits forming logical qudits. We require one of these qudits per dimension to represent the associated integer coefficient, with the number of qubits composing each qudit register depending on the integer range required and choice of encoding \cite{joseph2021two}. See Fig. \ref{fig:qubit_diag} for a schematic diagram of how the operators are associated to the qubit registers for which they decode to integers.
        Each circle in the diagram represents a qubit (with values here in the computational basis). Registers of qubits forming a qudit are organised by columns, with an operator $ \Q $ acting on the a particular column to decode the qubit values and produce an integer coefficient $ z_i $. The lines between sites represent nearest neighbour connections. For the QUBO setting a complete graph topology is required.
        
        The different encodings can be unified by formulating them as a map from a set of $ k $ qubits forming a qudit to an integer range,
        \begin{equation}\label{eq: QUBO Ising Ham}
            \Q_{enc}:\{-1,+1\}^k\to[-f(k),f(k)],
        \end{equation}
        where $ f(k) $ is dependent on the specific encoding. This formulation results in the corresponding Ising Hamiltonian of the form,
        \begin{equation}
            \hat{H} = \sum_{\langle i, j\rangle}\Q_{enc}^i\Q_{enc}^j\G_{ij},
        \end{equation}
        where the index on the operator $ \Q_{enc}^i $ refers to the operator acting on the $ i^{th} $ qudit, i.e. the $ i^{th} $ register of qubits. Here we compare three different integer encodings.
    
            \paragraph{Binary Encoding} The binary operator takes a register of qubits with states in the $ \{-1,+1\} $ basis and maps them to the computational basis from which it assigns a value of $ 2^k $ to the $ k^{th} $ qubit as in standard binary codes. The form of the operator is,
            \begin{align}\label{eq: bin encoding}
                &\Q_{bin}:\{-1,+1\}^k\to[-2^{k-1},2^{k-1}]; \\
                &\Q_{bin} = \sum_{p=0}^k 2^{p-1} \pZ_p + \frac{\I}{2},
            \end{align}
            where $ \pZ $ is the Pauli-Z operator corresponding to the $ p^{th} $ qubit of a qudit register. The final output is shifted down by $ 2^k $ to give an integer range of $ [-2^k, 2^k-1] $. Whilst this operator is the most efficient encoding of integers by a binary string it has the downside of not being particularly resistant to noise as it lacks any redundancy. Due to the exponential significance of each qubit within the encoding the binary encoding also leads to an exponential growth in the size of coupling constants in an Ising model implementation, as seen in \cite{joseph2021two}.
    
            \paragraph{Hamming Encoding} The Hamming encoding measures the Hamming weight of the code to obtain an integer by counting the number of qubits in the $ +1 $ state with form,
            \begin{align}\label{eq: ham encoding}
                &\Q_{ham}:\{-1,+1\}^{2^{k+1}}\to[-2^k, 2^k]; \\
                &\Q_{ham} = \frac{1}{2}\sum_{p=1}^{2^{k+1}}\pZ_p.
            \end{align}
            Note that in this case we require $ 2^k $ qubits. This is also shifted to be symmetric around zero with output $ [-2^k, 2^k] $. The Hamming encoding has inbuilt redundancy making it a more attractive choice for implementing on near term quantum architecture, however we lose the exponential scaling in the number of integers represented found in the binary case.
    
            \paragraph{Polynomial Encoding} The polynomial encoding seeks to find a mid ground between the information density, but exponential coupling constant growth, of the binary encoding, and the redundancy but low information density of the Hamming encoding. It takes the form,
            \begin{align}\label{eq: poly encoding}
                &\Q_{poly}:\{-1,+1\}^k\to[-m,m]; \\
                &\Q_{poly} = \frac{1}{2}\sum_{p=1}^k p\pZ_p + \frac{\I}{2}\left(\left\lceil\frac{k}{2}\right\rceil\mod2\right).
            \end{align}
            Here the second term is required to ensure an integer is always returned. This returns integer in the range $ [-m, m] $ where,
            \begin{equation}
                m = \frac{k^2 + k}{4}.
            \end{equation}
            This particular polynomial encoding is quadratic in nature, however it can be generalised for any polynomial by adjusting the exponent of the $ p $ term in Eq. \eqref{eq: poly encoding}. In this work we restrict ourselves to the quadratic case, referring to this as the polynomial encoding.

        In section \ref{sec:sims} these encodings - including the following adaptation - are used to apply a CIM to the SVP, with spatial complexity bounds on the number of qubits required to implement these algorithms presented here. 

    \subsection{Removing the Zero Vector\label{sec:0 restriction}}
        When computing solutions for a computational problem by encoding them into an Ising model one typically seeks to have the ground state respresent the solution to the problem. In the above formulation a possible state represents the integer vector $ \bm{0} $, corresponding to $ \bm{0}\in\L $. The length of this vector is naturally shorter than the first succesive minima, $ \lambda_1 $ - representing the length of the shortest non-zero vector - and therefore the state representing this will be the ground state. However by restricting the integer range that a qudit register can represent the ground state can instead be made to represent the shortest vector in the lattice.

        The proposed method is as follows: one qudit at a time  - i.e. taking the integer coordinate for one dimension at a time - restrict the integer range represented by the qudit to $ [1, m] $ for some chosen positive integer $ m $. This removes the zero vector from the search space. However, for a given basis it is not known whether none of the integer coordinates will be zero for the shortest vector, so this Ising solver in this case must be repeated $ n $ times for a lattice $ \L\subset\R^n $ to ensure that the shortest vector is apparent across the total set of vectors accessible by this method. Formally, we define a family of lattice subsets $ \{\tilde{\L}_i\}_{i\in[n]} $,
        \begin{equation}\label{eq: search sublattices}
            \tilde{\L}_i:=\B\Z^{i-1}\oplus\Z_{>0}\oplus\Z^{n-i},
        \end{equation}
        for which we apply the quantum algorithm to, repeating for each value of $ i\in[n] $. Here the notation $ \B\mathcal{A} $ denotes all combinations of the basis $ \B $ with elements of the set $ \mathcal{A} $, in this case a subset of the integers, $ \tilde{\Z}^n=\Z^{i-1}\oplus\Z_{>0}\oplus\Z^{n-i} $. In practice the quantum algorithm is applied over a subset bounded of an individual $ \tilde{\L}_i $ out of computational necessity, with the bounds discussed in \ref{bounds}.
        
        To implement this method new qudit operators are required, to be used in concert with the standard operators, and only applied to the qudit chosen to be restricted.

        \paragraph{Binary encoding} To restrict this range to \\ $ [1,2^k+1] $ we define the following operator,
        \begin{align}
            &\Q'_{bin}: \{-1,+1\}^r \to [1,2^k+1]; \\
            &\Q'_{bin}:= \sum_{p=0}^r2^{p-1}\pZ_p + \frac{2^{r+1}+1}{2}\I,
        \end{align}
        where $ r=k-1 $.

        \paragraph{Hamming encoding} To restrict this range to $ [1, 2^k+1] $ we define the following operator,
        \begin{align}
            &\Q'_{ham}:\{-1,+1\}^{2^k}\to[1,2^k+1]; \\
            &\Q'_{ham}:=\sum_{p=1}^{2^k}\frac{\pZ_p}{2} + (2^{k-1}+1)\I.
        \end{align}

        \paragraph{Polynomial Encoding} To restrict this range to $ [1,m+1] $ we define the following operator,
        \begin{align}
            \Q'_{poly}&:\{-1,+1\}^{r}\to[1,m+1]; \\
            \Q'_{poly}&:=\frac{1}{2}\sum_{p=1}^rp\pZ_p+\frac{r(r+1)+4}{4}\I \\
            &\quad+\frac{\I}{2}\left(\left\lceil\frac{r}{2}\right\rceil\mod2\right),
        \end{align}
        where,
        \begin{equation}
            r = \left\lceil\frac{\sqrt{2k(k+1)-7}-1}{2}\right\rceil.
        \end{equation}

        Whilst this requires the ground state of the Ising model to be computed a factor of $ n $ times over to ensure that the shortest vector is within the overall solution space this reformulation should sample the shortest vector with greater efficacy than the unrestricted formulation. It is incorporated into the numerical simulations in section \ref{sec:sims}.

        \subsubsection{Applying the Restriction to Cyclic and Negacyclic Lattices}
            If we let $ \L $ be an cyclic or negacyclic lattice this method of removing the zero vector as the ground state can be simplified, only having to restrict a single qudit register. This is due to the following lemma, that ensures the existence of a shortest vector for each lattice subset defined in Eq. \eqref{eq: search sublattices},
            \begin{lemma}\label{lem:well rounded}
                Let $ \L\subset\R^n $ be a well rounded lattice, with a set of minimal vectors,
                \begin{equation}
                    S(\L):=\{\v\in\L\mid\|\v\|=\lambda_1\},
                \end{equation}
                where $ \lambda_1 $ is the first successive minima of $ \L $. Fixing a basis for the lattice, $ \B $, such that $ \L=\L(\B\Z^n) $, define a family of lattice subsets $ \{\tilde{\L}_i\}_{i\in[n]} $,
                \begin{equation}
                    \tilde{\L}_i:=\B\Z^{i-1}\oplus\Z_{>0}\oplus\Z^{n-i}.
                \end{equation}
                Then $ \tilde{\L}_i\cap S(\L)\neq\emptyset $ for all $ i\in[n] $.
                \begin{proof}
                    See appendix \ref{app:proofs}.
                \end{proof}
            \end{lemma}
            In other words there is always intersection between the set of minimal vectors - solutions to SVP - and each lattice subset used in the restricted formulation. Since every lattice subset this method searches over is guaranteed to contain a solution to shortest vector problem for cyclic or negacyclic lattices one can pick an arbitrary $ i\in[n] $ as the restricted dimension. This reduces the time complexity of implementing this restriction by a factor of $ n $, bringing it in line with the unrestricted approach. There will be a particular choice of $ i $ such that the integer coefficients are minimised, also increasing efficacy of the quantum algorithm. However, the disparity between the size of the integer coefficients for each solution can be minimised by applying a basis reduction prior to the quantum algorithm. 

    \subsection{\label{bounds}Spatial Complexity Bounds}
        A key question one can ask about this framework is how does the number of qubits required to ensure the shortest vector is within the search space scale with the dimension of the problem? To answer this first requires a bound on the integer coefficients for a lattice vector, provided Lem. 1 of \cite{albrecht2022variational}, stated here without proof.
        \begin{lemma}[\cite{albrecht2022variational}]\label{lem:albrect}
            Let $ \L=\L(\B) $ be a full rank lattice with basis $ \B $ and let $ x_1,\ldots,\x_n \in\Z $ be such that $ \|x_1\cdot\b_1+\cdots+x_n\cdot\b_n\|\leq A $. then for all $ i = 1,\ldots, n $ we have $ |x_i|\leq A\|\b^*_i\| $ where $ \b_1^*,\ldots,\b^*_n $ are rows of the dual basis $ \B^* $.
        \end{lemma}
        The choice of the search radius $ A $ such that $ \lambda_1\leq A $ but $ A $ is not too large is an important consideration. For random lattices the Gaussian Heuristic \cite{chen2016measure} provides a good estimate as,
        \begin{equation}
            gh(\L) := \sqrt{\frac{n}{2\pi e}}\vol(\L)^{1/n}.
        \end{equation}

            By setting $ A=gh(\L) $ one can use Lem. \ref{lem:albrect} to bound the number of qubits required to ensure the existence of the shortest vector within the search space. For this work we restrict ourselves to cyclic and negacyclic lattices to provide spatial complexity bounds for all three encodings, in the case of an HKZ reduced basis. Proofs of the following theorems can be found in appendix \ref{app:proofs}. 
            \begin{theorem}\label{thm: bin qubit ideal}
                Let $ \L\subseteq\Z^n $ be a full rank, cylcic or negacyclic lattice such that $ \L=\L(\B) $, where $ \B $ is an HKZ reduced basis. Then there exists a quantum algorithm employing the binary qubit encoding requiring,
                \begin{equation}
                    Q_{bin}(n)\leq\frac{3n}{2}\log_2(n+3)-\frac{n}{2}(2+\log_2\pi e)+\O(1),
                \end{equation}
                qubits to ensure the existence of the shortest vector within the search space.
            \end{theorem} 
            
            \begin{theorem}\label{thm: ham qubit ideal}
                Let $ \L\subseteq\Z^n $ be a full rank, cyclic or negacyclic lattice such that $ \L=\L(\B) $, where $ \B $ is an HKZ reduced basis. Then there exists a quantum algorithm employing the Hamming qubit encoding requiring,
                \begin{equation}
                    Q_{ham}(n)\leq2^{\O(1)}n^{5/2}\approx\O(n^{5/2}),
                \end{equation}
                qubits to ensure the existence of the shortest vector within the search space.
            \end{theorem}

            \begin{theorem}\label{thm: poly qubit ideal}
                Let $ \L\subseteq\Z^n $ be a full rank, cyclic or negacyclic lattice such that $ \L=\L(\B) $, where $ \B $ is an HKZ reduced basis. Then there exists a quantum algorithm employing the polynomial qubit enocoding requiring,
                \begin{equation}
                    Q_{poly}(n)\leq2^{O(1)}n^{7/4}\approx\O(n^{7/4}),
                \end{equation}
                qubits to ensure the existence of the shortest vector within the search space.
            \end{theorem}

            \paragraph{Comparing Bounds}
                \begin{figure}
                    \centering
                    \includegraphics[scale=0.425]{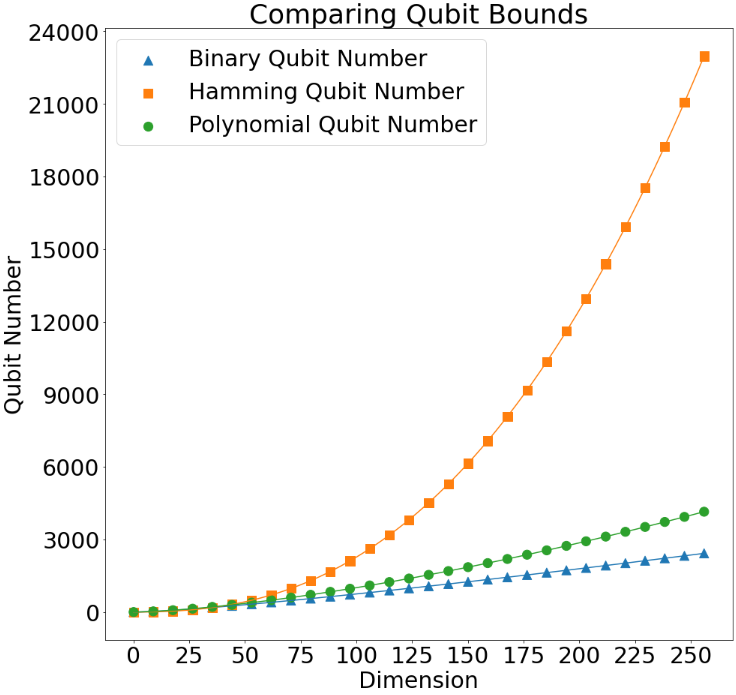}
                    \caption{A numerical comparison of the bounds from Thm. \ref{thm: bin qubit ideal}, \ref{thm: ham qubit ideal} and \ref{thm: poly qubit ideal} against the dimension of the lattice. The binary qubit number can be seen in the blue triangle curve, Hamming orange square curve and polynomial green circle curve. As the the dimension scales the binary encoding remains the most spatially efficient, however the polynomial encoding is not far behind lending credence to its use as it also overcomes several of the issues of the binary encoding.}
                    \label{fig:comparing bounds}
                \end{figure}
                Fig. \ref{fig:comparing bounds} plots a numerical comparison of the bounds found in Thm. \ref{thm: bin qubit ideal}, \ref{thm: ham qubit ideal} and \ref{thm: poly qubit ideal}. As expected - given the growth of the respective state spaces - the binary is the most spatially efficient, with the Hamming the least efficient and the polynomial inbetween the two. The polynomial bound is in fact only $ \O(n^{7/4}-n\log_2n) $ worse than the binary, which alongside its ability to overcome several of the issues found in the binary case makes it an attractive candidate for QUBO SVP algorithms.

\section{\label{sec:sims}A CIM SVP Solver}
    We compare the efficacy of the CIM with CFC error correction as a SVP solver using the QUBO formulation in simulation for an initial lower dimensional setting and the above restriction of $ \bm{0}\notin\L $.

    To implement the QUBO Ising model stated in Eq. \eqref{eq: QUBO Ising Ham} with the above operators requires single operator terms, which correspond to external field considerations in a physical Ising model - a feature the CIM can not implement. Instead the above system is replaced by a system containing an auxiliary site interacting with all other sites, for which the state is kept constant. Operators in this system take the form,
    \begin{align}
        \Q_{bin}^{(i)} &= \sum_{p=0}^{k}2^{p-2}\pZ_{ip} + 2^{-1}\pZ_A \\
        \Q_{poly}^{(i)} &= \sum_{p=1}^kp\pZ_{ip} + \frac{\pZ_A}{2}\left(\left\lceil\frac{k}{2}\right\rceil\mod2\right),
    \end{align}
    where $ \pZ_A $ is the Pauli-$ z $ operator acting on the auxiliary qubit.

    \begin{figure*}
            \centering
            \includegraphics[scale=0.16]{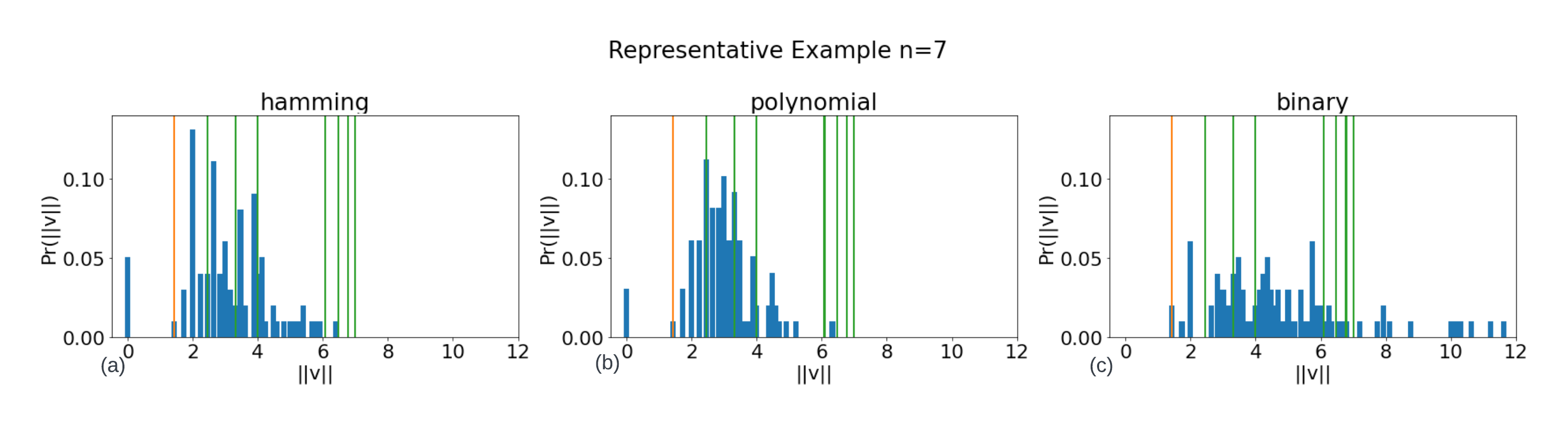}
            \caption{Sample output distribution for the CIM-CFC applied to the QUBO approach to the SVP for the three encodings, the Hamming (Eq. \eqref{eq: ham encoding}), polynomial (Eq. \eqref{eq: poly encoding}), and binary encodings (Eq. \eqref{eq: bin encoding}), for the lattice in dimension 7 defined by the basis given in Eq. \eqref{eq: CIM rep basis}. Here the orange line represents the length of the shortest vector in the lattice and the green lines represent the lengths of the input basis vectors.}
            \label{fig:CIM rep example}
        \end{figure*}
    
    \subsection{Initial Findings}
        In the low dimensional regime lattices are spanned by the rows of random integer matrices containing elements from $ \{-1, 0, 1\} $ - ensuring the existence of the the shortest vector within the basis. Bases are subsequently transformed by unimodular matrices with elements no greater than $ 6 $ (for computational simplicity) to create a problem set of hard bases. 
    
        Typically when running a CIM the parameters are varied over the course of the process. This is to encourage convergence to a state whilst still allowing for adequate exploration of the state space. Table \ref{tab:CIM parameters} contains a list of the parameters used in the lower dimensional simulations including how they varied. 
    
        \begin{table}[h]
            \centering
            \begin{tabular}{c||c}
                Parameter & Value  \\
                \hline
                \hline
                $ \Delta t $ & 0.0001 \\
                Time-steps & 320000 \\
                Varying Parameter Time Steps & 288000 \\
                Stationary Parameter Time Steps & 3200 \\
                $ \beta $ & 0.8 \\
                Initial $ p $ & -2 \\
                Stationary $ p $ & 0 \\
                Initial $ \alpha $ & 1 \\
                Stationary $ \alpha $ & 2
            \end{tabular}
            \caption{Simulation parameters for the CIM-CFC applied to the SVP. Varied parameters were varied smoothly over the varying time steps and then held at their stationary value until the end of the simulation.}
            \label{tab:CIM parameters}
        \end{table}
    
        The test set of low dimensional lattices were generated with entries from $ \{-1,0,1\} $, making sure that the shortest vector was already apparent within the basis, then augmented by unimodular matrices with elements no greater than 6 for computational simplicity.  By considering each run as sampling from the lattice in particular we can assess the accuracy of the CIM-CFC as an Ising solver for this problem by examining the output sample distribution. 
    
        % \begin{figure*}
        %     \centering
        %     \includegraphics[scale=0.3]{figures/CIM_rep_example.png}
        %     \caption{Sample distribution for the CIM-CFC applied to the QUBO approach to the SVP for all three encodings in dimension 7 with basis \eqref{eq: CIM rep basis}. Here the orange line represents the length of the shortest vector in the lattice and the green lines represent the lengths of the input basis vectors.}
        %     \label{fig:CIM rep example}
        % \end{figure*}
        \subsubsection{Representative Example}
            In Fig. \ref{fig:CIM rep example}, a representative example lattice for all three qubit encodings - hamming, polynomial and binary - in dimension 7, for a lattice spanned by the rows of,
            \begin{equation}\label{eq: CIM rep basis}
                \B = \begin{pmatrix}
                    2 & 3 & -1 & -3 & 3 & 1 & 4 \\ 
                    1 & 2 & -1 & -2 & 2 & 1 & 1 \\
                    0 & 0 & -3 & -4 & 4 & 0 & 1 \\
                    -1 & -2 & -1 & -2 & 1 & 0 & 0 \\ 
                    -1 & -2 & 3 & 4 & -4 & 0 & 0 \\
                    1 & 1 & -1 & -1 & 1 & 0 & 1 \\
                    1 & 3 & 3 & 2 & -2 & 3 & 1
                \end{pmatrix},
            \end{equation}
            can be seen. The green lines represent the lengths of the basis vectors and the orange line (the furtherest left line in each plot) the length of the shortest vector. In this example output sample, the CIM SVP solver does sample the shortest vector with non-zero probability, as well as finding the ground state $ \bm{0}\in\L $. However, despite the fact that it samples the shortest vector much less frequently than large vectors and the zero vector the distribution also skews shorter than the basis vectors suggesting that this could be used to solve the approximate short vector problem. Whilst this sample distribution is skewed toward shorter vectors the Ising system fails to achieve the ground state more often than excited states, and for the binary encoding does not find the ground state at all. Considering this purely from the perspective of an Ising solver this failure to find the ground state presents a problem, however as an SVP solver the ground state under this formulation represents the trivial solution, as such this imperfect operation can be considered useful. We can see from comparing SubFig. \ref{fig:CIM rep example}(a) and \ref{fig:CIM rep example}(b) to SubFig. \ref{fig:CIM rep example}(c), the Hamming and polynomial encodings are more effective in sampling short lattice vectors in this instance.

            \begin{figure*}
                \centering
                \includegraphics[scale=0.16]{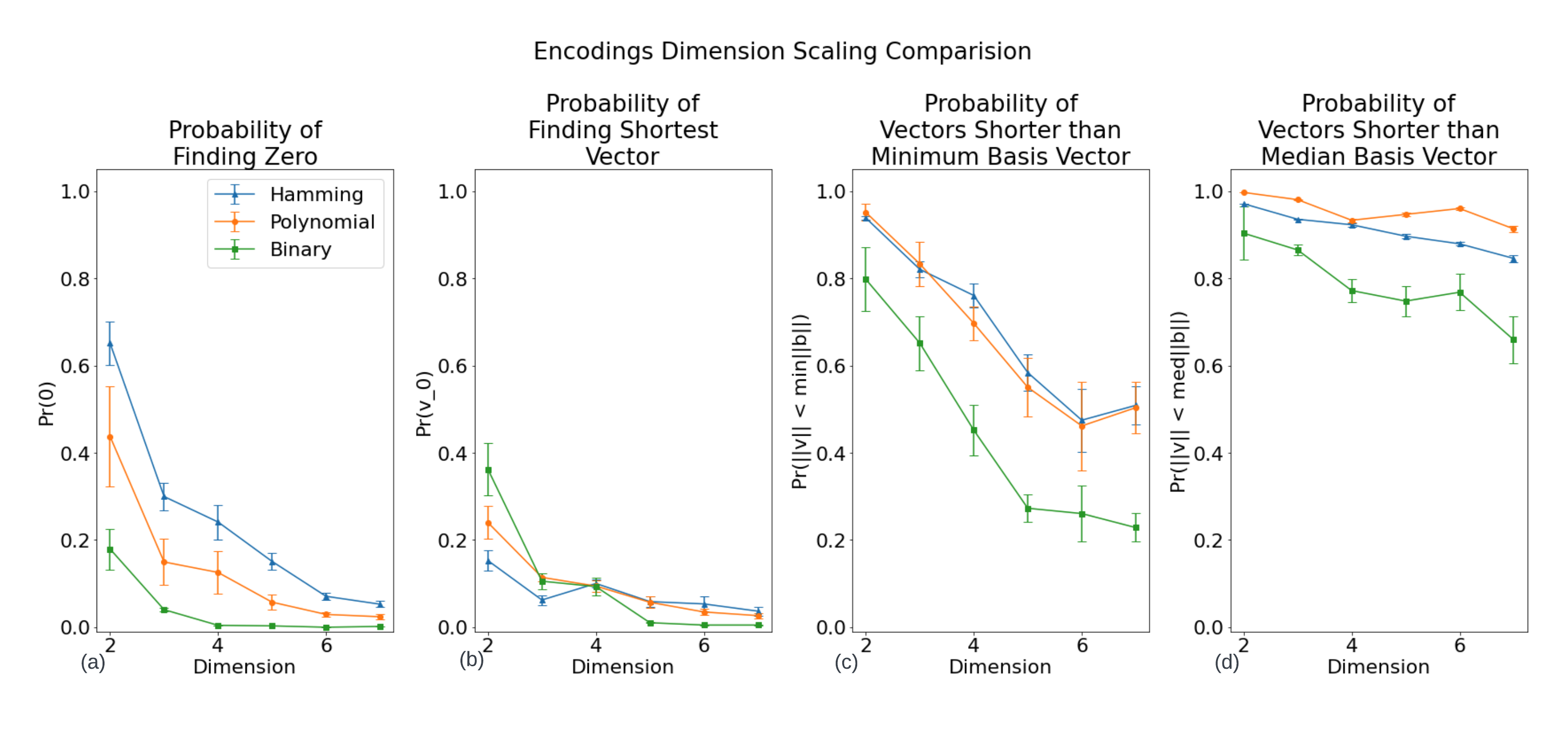}
                \caption{Comparing the efficacy of the CIM-SVP solver across dimensions 2-7 for the all three encodings, the Hamming (blue triangle), polynomial (orange circle), and binary (green square) encodings. Here the four panels correspond to the four figures of merit stated above: the probability of measuring the zero vector (Eq. \eqref{eq:FOM zero}), the probability of measuring the solution to SVP (Eq. \eqref{eq:FOM svp}), the probability of measuring vectors shorter than the minimum basis vector (Eq. \eqref{eq:FOM min basis}) and the probability of measuring vectors shorter than the median basis vector (Eq. \eqref{eq:FOM med basis}). The results are aggregated over all lattices in the set of test bases, with the mean value for each figure of merit given per dimension.}
                \label{fig: standard run}
            \end{figure*}
        \subsubsection{Aggregated Results}
            To understand how this algorithm performs the dimension of the problem is scaled averaged over all the lattices, with four key figures of merit (FoM) comparing the efficacy of the three encoding variants, for sampled lattice vectors $ \v\in\L=\L(\B) $ for a given input basis $ \B $,
            \begin{enumerate}
                \item The mean sample probability of the zero vector,
                \begin{equation}\label{eq:FOM zero}
                    \Pr(\|\v\|=0).
                \end{equation}
                \item The mean sample probability of the shortest vector,
                \begin{equation}\label{eq:FOM svp}
                    \Pr(\|\v\|=\lambda_1),
                \end{equation}
                where $ \lambda_1 $ is the first successive minima.
                \item The mean sample probability of a vector being shorter than the minimum basis vector,
                \begin{equation}\label{eq:FOM min basis}
                    \Pr(\|\v\|\leq\min_{i\in[n]}\|\b_i\|),
                \end{equation}
                for $ \b_i\in\B $.
                \item The mean sample probability of a vector being shorter than the median basis vector,
                \begin{equation}\label{eq:FOM med basis}
                    \Pr(\|v\|\leq\text{med}\|\b_i\|),
                \end{equation}
                for $ \b_i\in\B $.
            \end{enumerate}
            \begin{figure*}
                \centering
                \includegraphics[scale=0.16]{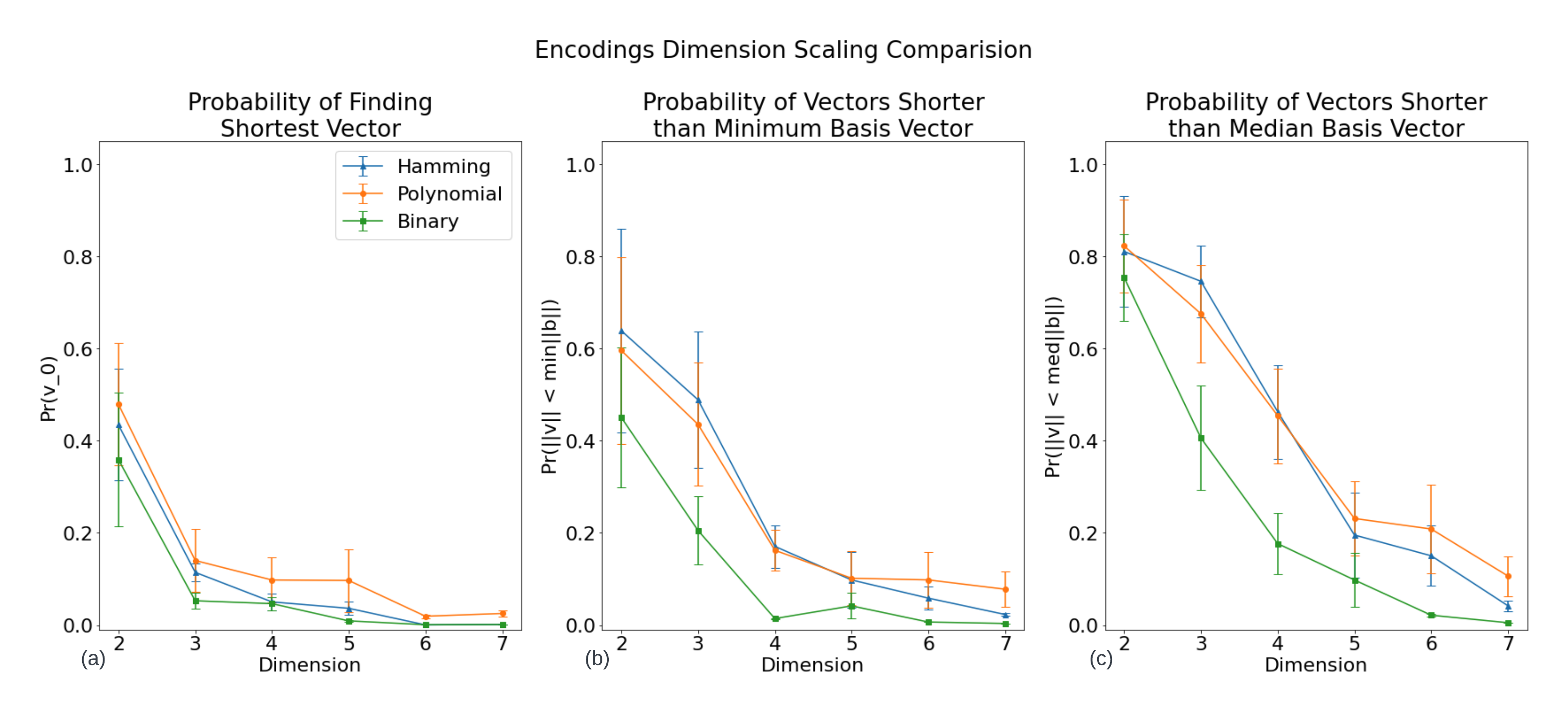}
                \caption{Comparing the efficacy of the CIM-SVP solver with the condition $ \bf{0}\notin\L $ across dimensions 2-7 for the all three encodings, the Hamming (blue triangle), polynomial (orange circle), and binary (green square) encodings. Here the three panels correspond to three of the four figures of merit stated above: the probability of measuring the solution to SVP (Eq. \eqref{eq:FOM svp}), the probability of measuring vectors shorter than the minimum basis vector (Eq. \eqref{eq:FOM min basis}) and the probability of measuring vectors shorter than the median basis vector (Eq. \eqref{eq:FOM med basis}). The results are aggregated over all lattices in the set of test bases, with the mean value for each figure of merit given per dimension.}
                \label{fig:CIM res dim hamming box plots}
            \end{figure*}
            Fig. \ref{fig: standard run} aggregates the results of 1000 CIM runs with the above parameters, for each of the 32 test lattices in dimensions $ n=\{2,3,4,5,6,7\} $. All three algorithms exhibit a decreasing success in each of the four FoMs as the lattice dimensions is increased, a result that is to be expected given the increases in complexity of these systems as the dimension scales. In general the Hamming (blue triangle) and polynomial (orange circle) results sit above the binary (green square), indicating a higher probability of success for all metrics in each dimensions for these two encodings. 

            The first metric, the probability of the CIM system finding the ground state (representing the zero vector in this case) - given by Eq. \eqref{eq:FOM zero} - is compared for each encoding in SubFig. \ref{fig: standard run}(a). This is a good indicator in the successful operation of the CIM, with the Hamming encoding showing the greatest success, followed by the polynomial then binary. This ordering of success, however, is likely due to the degeneracies in integer representation apparent in each encoding. Within in the binary encoding each integer - and therefore each lattice vector - is uniquely represented, a property not preserved in the other two encodings. Both the Hamming and polynomial encodings over represent small integers, with a Gassian like distribution over the integers centered at zero with a wide variance. 

            In contrast to SubFig. \ref{fig: standard run}(a), the separation in successful operation of the three encodings in finding the shortest vector of the lattice, given by SubFig. \ref{fig: standard run}(b), is not as clear. The binary encoding has greater success initially, likely due to the lack of degeneracies in integer representation. However as the dimensions scale the Hamming and polynomial encodings are still able to produce short vectors, the binary encoding failing to compute the shortest vector in the overwhelming majority of cases. This figure of merit is of the greatest consequence as it measures the efficacy in directly solving SVP.

            SubFig. \ref{fig: standard run}(c) and \ref{fig: standard run}(d) show the figures of merit corresponding to the probability of the vectors returned having shorter lengths than the shortest and median input basis vectors respectively. This allows us an insight into how effective this algorithm is at solving the approximate shortest vector problem - useful for both the cryptanlysis of schemes based on the approximate form of the problem \cite{nguyen2006cryptanalysis}, and in producing short vector candidates for basis reductions algorithms \cite{zhu2022iterative}. Here the FoM shows a high probability of success for all three encodings, notably with a far less steep drop off in this efficacy as the lattice dimension is scaled. 

            A key takeaway from figure \ref{fig: standard run} is the success of the Hamming and polynomial encodings over the binary in all metrics, especially the polynomial given its relatively similar complexity to the binary, whilst simultaneously solving several of the extant problems inherent in the binary encoding - for instance the exponential scaling in the coupling coefficients. What this means for the security for lattice based cryptosystems is less clear cut, with the decay of the probability of finding the shortest vector being the primary indicator. To assess this simulations in dimensions not currently computationally accessible, or implementations on physical CIMs would be required. This would also necessitate a further investigation into optimising the parameters of the CIM, which is beyond the scope of this work.

    \subsection{Removal of Zero Vector Restriction}
        So far the method for solving the shortest vector problem has been to apply the CIM in an atypical way, with the assumption that the CIM is not a perfect Ising solver. In this way we presume that the ground state is not going to be the final state the system settles in. The two degenerate first excited states in this formulation represents the shortest non zero vector and the negative shortest vector, with each subsequent excited state representing the successive minima. As seen above this allows the CIM implementation to have success in the approximate shortest vector problem. However by restricting the integer range that a qudit can represent the ground state can instead be made to represent the shortest vector in the lattice. This will become an especially important technique as the efficacy of Ising solvers in application to this problem increases.
    
        To implement this restriction the QUBO system is adapted as in section \ref{sec:0 restriction}. The alternative qudit operator is applied to one dimension at a time producing an integer $ z\in[1,f(k)] $, where $ f(k) $ depends on the choice of encoding. This is then repeated for each dimension to ensure that the shortest vector is within the total search space - adding a multiplier to the time complexity of $ n $. As above the system is adapted to include an auxiliary site to replicate external field considerations.

        As in Fig. \ref{fig: standard run}, Fig. \ref{fig:CIM res dim hamming box plots} plots the results of repeated simulation of the CIM across dimensions 2-7, aggregated across the 32 test lattices per dimension, in this case for the restricted formulation. Only the figures of merit giving the probability of sampling the shortest vector and vectors shorter than the minimum and median basis vectors are shown as the zero vector is no longer a valid solution under this formulation. The restricted dimension that encodes the shortest vector has been post-selected for in these results, meaning that if all the samples for each repeated case were collated the results in SubFig. \ref{fig:CIM res dim hamming box plots}(a) would be scaled down by approximately a factor of $ n $, however we would expect to see an increase in the probabilities found in SubFig. \ref{fig:CIM res dim hamming box plots}(b) and \ref{fig:CIM res dim hamming box plots}(c). Similar to the unresticted case given in Fig. \ref{fig: standard run}, the Hamming (blue triangle) and polynomial (orange circle) encodings out perform the binary (green square) encoding across all three figures of merit, with a decrease in efficacy across all three as the lattice dimension increases. However, this decrease is sharper than in the unrestricted case, especially in the sampling of short vectors with respect to the basis. 

        In addition to this, whilst we no longer see any occurrences of the trivial solution to SVP, $ \bm{0}\in\L $, we do not observe a great improvement in the ability of the restricted system to directly solve SVP. This is likely to due to the energy landscape of the search space becoming more jagged - that is to say not smooth - than the above method, with an emergence of local minima representing vectors of greater length. Also contributing is the fact that by restricting the output to only positive integers only the positive shortest vector is within the search space, similarly only positive vectors representing the successive minima are evident. 

\section{\label{sec:conc}Discussion}
    The CIM SVP solver presented here shows promise in solving SVP on average for the test lattices considered. When considering the question of scaling this problem there is a drop off in the ability of such a system in solving exact SVP, however it could still aid in the solution of approximate SVP. The efficacy of the system presented in this work can certainly be improved upon by optimising the parameters of the CIM, especially as higher dimensions are considered. CIMs explore the state space most effectively when the system is kept around a phase change, though uncovering the phase changes of an Ising model is a hard problem in and of itself. Another method to increase the efficacy of the above CIM SVP solver would be to explore alternative encodings. This would provide an alternative energy landscape over the state space, which if carefully chosen could provide better optimisation. Examples of these encodings include Gray code ordering \cite{sawaya2020resource}, which would make the energy shift required for a change of one integer in a binary representation uniform as each successive bitstring in Gray code ordering varies by one bit. However, for a $ k $-length bitstring this requires $ k $-local interactions, which are not attainable by the CIM (and require $ 2^{k-1} $ CNOT gates to encode in a gate model).
    
    Physical CIMs have already seen success with implementing large numbers of qubits, and with the advent of error correction their application to lattice problems within the QUBO formulation in a higher dimensional setting could see success if the issues discussed above can be overcome. This could also be well placed within MIMO fields as a decoder.

    Analysis of the algorithms in section \ref{sec:QUBO} reveal that the binary encoding will continue to prove viable as the dimension of the lattice problems is scaled to cryptographically relevant dimensions. However, the scaling in the coupling strengths required for the binary encoding discloses a large disparity between interactions of the least significant qubits and the rest of the system, a feature not apparent in the Hamming and polynomial encodings. To implement the binary encoding therefore requires a sharp scaling on the interaction coefficients, leading to the potential for greater error propagation. The polynomial encoding provides a solution to these problems by introducing a less sharp scaling in the coefficients, at the cost of some efficiency. Along with a the numerical results above, this suggests that the polynomial encoding would provide a suitable alternative to the binary encoding. However, due to the redundancy in the representation of integers in both the Hamming and polynomial encodings, these over represent small integers - with a zero centered, wide variance, Gaussian like distribution over $ \Z $. If the integers required for short vectors in a particular basis are found at the extremes of the range represented both the Hamming and polynomial encodings could have difficulty in finding these vectors.

\begin{acknowledgments}
    We would like to thank Andrew Mendhelson, Sam Reifenstein, Satoshi Kako and Yoshihisa Yamamoto for their helpful discussions. This work was supported in part by the Engineering and Physical Sciences Research Council (EPSRC) under Grant No. EP/S021043/1.
\end{acknowledgments}

% \nocite{*}
% \newpage
\bibliography{bib}% Produces the bibliography via BibTeX.

% \clearpage
\appendix
\section{\label{app:proofs}Proof of Various Statements}
    Collected proofs of various statements made above.

    \subsection{Proof of Lem. \ref{lem:well rounded}}
        The proof of Lem. \ref{lem:well rounded} relies on the following technical lemma,
        \begin{lemma}\label{lem:tech lemma}
            Let $ A\in\R^{n\times n} $ be a full rank matrix. Then there exists a row permutation $ P $ such that $ PA $ has a zero free diagonal, i.e. $ (PA)_{ii}\neq0 $ for all $ i\in[n] $.
            \begin{proof}
                $ A=(a_{ij}) $ is full rank, and therefore non-singular: $ \det(A)\neq0 $. Consider the following definition of the determinant,
                \begin{equation}
                    \det(A) = \sum_{\sigma\in S_n}\left(\text{sign}(\sigma)\prod_{i=1}^n a_{i\sigma_i}\right).
                \end{equation}
                If there does not exists a permutation $ P $ such that $ PA $ has a zero-free diagonal then $ \prod_{i=1}^n a_{i\sigma_i}=0 $ for all $ \sigma\in S_n $, which would set $ \det(A)=0 $. Thus we find a contradiction and there must exist such a $ P $.
            \end{proof}
        \end{lemma}
        Now we provide the proof of Lem. \ref{lem:well rounded}:
        \begin{proof}
           By the well rounded nature of $ \L $, $ \text{Sp}_{\R}(S(\L))=\R^n $. Then we can construct a linearly independent subset $ \tilde{S}(\L)=\{\v_1,\ldots,\v_n\}\subset S(\L) $. This defines the matrix $ S=[\v_1:\cdots:\v_n]^T $, which is related to the basis by,
            \begin{equation}
                S = NB,
            \end{equation}
            where $ B=[\bm{b}_1:\cdots:\bm{b}_n]^T $ is the basis matrix, and $ N=[\bm{z}_1:\cdots:\bm{z}_n]^T $ the matrix of integer coefficients relating $ S $ to $ B $. The above condition is equivalent to exhibiting a manipulation of $ N $ such that $ N_{ii}>0 $ for all $ i\in[n] $.
            
            Since $ S $ and $ B $ are full rank, $ N $ is also full rank. Then the following facts allow us to construct $ N $ that fulfills the above condition:
            \begin{enumerate}
                \item By Lem. \ref{lem:tech lemma} there exists a row permutation matrix $ P $ such that $ N'=PN $ and $ N'_{ii}\neq0 $. This is applied as $ PS=PNB $, which also reorders the rows of $ S $ accordingly, but retains the linear independence and minimality of the vectors.
                \item For each $ \v_i\in\tilde{S}(\L) $ if the corresponding integer value $ N_{ii}<0 $ we can replace $ \v_i\to-\v_i $. This can be done as for each $ \v\in S(\L) $ $ -\v\in S(\L) $, for which the substitution into $ \tilde{S}(\L) $ maintains the linear independence of the set.
            \end{enumerate}
            The final set will fulfill the above constraints. Since this set can always be constructed each lattice subset $ \tilde{\L}_i $ must contain at least one minimal vector.
        \end{proof}

    \subsection{Proofs of Thm. \ref{thm: bin qubit ideal}-\ref{thm: poly qubit ideal}}
        For the proofs of these theorems we take advantage of the well rounded nature of cyclic lattices by way of the relation,
        \begin{equation}
            \lambda_i(\L)=\lambda_1(\L),
        \end{equation}
        for all $ i\in[n] $. Further, if we have an HKZ reduced basis $ \B=\{\b_1,\ldots,\b_n\} $ for a lattice $ \L $,
            \begin{equation}
                \|\b_i\|\leq\frac{\sqrt{i+3}}{2}\lambda_i(\L),
            \end{equation}
            for the $ i^{th} $ successive minima of the lattice $ \lambda_i(\L) $.
        \paragraph{Proof of Thm. \ref{thm: bin qubit ideal}}
            \begin{proof}
                Let $ k_i $ be the number of qubits in register $ i $, and $ m_i=\max|z_i| $ where $ z_i $ is the integer represented by the qubit register. For the binary encoding this is related as $ k_i=\lfloor\log_2(2m_i)\rfloor+1 $. Then,
                \begin{widetext}
                    \begin{align*}
                        Q_{bin}(n) &= \sum_{i=1}^n(\lfloor\log_2(2m_i)\rfloor+1), \\
                        &\leq2n + \sum_{i=1}^n\log_2m_i, \\
                        &\leq2n + \frac{n}{2}\log_2\left(\frac{n}{2\pi e}\right) + \sum_{i=1}^n\log_2\frac{\|\b^*_i\|}{\vol(\L^*)^{1/n}}, \\
                        &\leq\frac{n}{2}(1-\log_2\pi e)+\frac{n}{2}\log_2n\sum_{i=1}^n\log_2\frac{\sqrt{i+3}\lambda_i(\L^*)}{\vol(\L^*)^{1/n}}, \\
                        &\leq n\log_2(n+3)+\frac{n}{2}(1-\log_2\pi e)+\sum_{i=1}^n\log_2\frac{\lambda_1(\L^*)}{\vol(\L^*)^{1/n}}, \\
                        &\leq n\log_2(n+3)+\frac{n}{2}(1-\log_2\pi e)+\frac{n}{2}\log_2\gamma_n, \\
                        &\leq n\log_2(n+3)+\frac{n}{2}(1-\log_2\pi e)+\frac{n}{2}\log_2\left(\frac{1}{8}n+\frac{6}{5}\right), \\
                        &\leq\frac{3n}{2}\log_2(n+3)-\frac{n}{2}(2+\log_2\pi e)+\O(1).
                    \end{align*}
                \end{widetext}
                Giving the desired form.
            \end{proof}

        \paragraph{Proof of Thm. \ref{thm: ham qubit ideal}}
            \begin{proof}
                Let $ k_i $ be the number of qubit sin register $ i $, and $ m_i=\max|z_i| $ where $ z_i $ is the integer represented by the qubit register. For the Hamming encoding this is related as $ 2k_i=m_i $. Then,
                % \begin{widetext}
                    \begin{align*}
                        Q_{ham}(n) &= \frac{1}{2}\sum_{i=1}^n m_i, \\
                        &\leq\sqrt{\frac{n}{8\pi e}}\sum_{i=1}^n\frac{\|\b^*_i\|}{\vol(\B^*)^{1/n}}, \\
                        &\leq\sqrt{\frac{n}{8\pi e}}\sum_{i=1}^n\frac{\sqrt{i+3}}{2}\cdot\frac{\lambda_i(\L^*)}{\vol(\L^*)^{1/n}}, \\
                        &=\sqrt{\frac{n}{32\pi e}}\sum_{i=1}^n\sqrt{i+3}\cdot\frac{\lambda_1(\L^*)}{\vol(\L^*)^{1/n}}, \\
                        &\leq\sqrt{\frac{n\gamma_n}{32\pi e}}\sum_{i=1}^n\sqrt{i+3}, \\
                        &\leq\sqrt{\frac{n\gamma_n}{32\pi e}}n\sqrt{n+3}.
                    \end{align*}
                % \end{widetext}
                By taking $ \log_2 $ of both sides we have,
                \begin{align*}
                    \log_2 Q_{ham}(n) & \leq2\log_2(n+3)+\frac{1}{2}\log_2\left(\frac{1}{8}n+\frac{6}{5}\right), \\
                    &\leq\frac{5}{2}\log_2n + \O(1),
                \end{align*}
                therefore,
                \begin{equation*}
                    Q_{ham}(n)\leq2^{\frac{5}{2}\log_2 n+\O(1)}\approx\O(n^{5/2}).
                \end{equation*}
            \end{proof}

    % \newpage
        \paragraph{Proof of Thm. \ref{thm: poly qubit ideal}}
            \begin{proof}
                Let $ k_i $ be the number of qubits in register $ i $, and $ m_i=\max|z_i| $ where $ z_i $ is the integer represented by the qubit register. First we derive a closed form for $ k $ in terms of $ m $,
                % \begin{widetext}
                    \begin{align*}
                        m &= \frac{1}{2}\sum_{i=1}^k i=\frac{k(k+1)}{4} \quad \implies \quad k=\frac{\sqrt{16m+1}-1}{2}.
                    \end{align*}
                % \end{widetext}
                Then,
                % \begin{widetext}
                \begin{align*}
                    Q_{poly}(n) &= \frac{1}{2}\sum_{i=1}^n(\sqrt{16m_i+1}-1), \\
                    &\leq\frac{5}{2}\sum_{i=1}^n\sqrt{m_i}-\frac{n}{2}, \\
                    &\leq \frac{5}{2}\left(\frac{n}{2\pi e}\right)^{1/4}\sum_{i=1}^n\sqrt{\frac{\|\b^*_i\|}{\vol(\L^*)^{1/n}}}-\frac{n}{2}, \\
                    &\leq \frac{5\sqrt{2}}{2}\left(\frac{n}{2\pi e}\right)^{1/4}\sum_{i=1}^n\sqrt{\frac{\sqrt{i+3}\lambda_i(\L^*)}{\vol(\L^*)^{1/n}}}-\frac{n}{2}, \\
                    &=\frac{5\sqrt{2}}{2}\left(\frac{n}{2\pi e}\right)^{1/4}\sum_{i=1}^n\sqrt{\frac{\sqrt{i+3}\lambda_i(\L^*)}{\vol(\L^*)^{1/n}}}-\frac{n}{2}, \\
                    &\leq\frac{5\sqrt{2}}{2}\left(\frac{n\gamma_n}{2\pi e}\right)^{1/4}\sum_{i=1}^n(i+3)^{1/4}-\frac{n}{2}, \\
                    &\leq\frac{5\sqrt{2}}{2}\left(\frac{n\gamma_n}{2\pi e}\right)^{1/4}n(n+3)^{1/4}-\frac{n}{2}.
                \end{align*}
                
                By taking $ \log_2 $ of both sides we find,
                \begin{align*}
                    \log_2 Q_{poly}(n) &\leq \frac{3}{2}\log_2 n+\frac{1}{4}\left(\frac{1}{8}n+\frac{6}{5}\right), \\
                    &\leq \frac{7}{4}\log_2 n + \O(1),
                \end{align*}
                therefore,
                \begin{equation*}
                    Q_{poly}(n)\leq2^{\frac{7}{4}\log_2n+\O(1)}\approx\O(n^{7/4}).
                \end{equation*}
            \end{proof}
            % \end{widetext}

\end{document}